\documentstyle[12pt,epsf]{article} 
 \newcommand {\bi} {\bibitem}
 \newcommand {\be} {\begin{equation}}
\newcommand {\bea} {\begin{eqnarray} \nonumber }
\newcommand {\ee} {\end{equation}}
\newcommand {\eea} {\end{eqnarray}}
 \newcommand {\eps} {\epsilon}
 \newcommand {\si} {\sigma}

\newcommand {\la} {\lambda}

 \newcommand {\al} {\alpha}

\newcommand {\lan} {\langle}
\newcommand {\ran} {\rangle}

\newcommand {\cN}  {{\cal N}}

\newcommand {\for} {\ \ \ \mbox{for}\ \ }

\newcommand {\cpa} {\right)}
\newcommand {\apa} {\left(}

\def \form#1 {eq. (\ref{#1}) }
\def \parziale#1#2  {{\partial {#1} \over \partial {#2}}}

\begin{document}   

\title{New ideas in glass transitions}
\author{  Giorgio Parisi \\ 
Dipartimento di Fisica,
Universit\`a {\sl La  Sapienza}\\ INFN Sezione di Roma I \\ Piazzale 
Aldo Moro, Roma 00187}
\maketitle
\begin{abstract}
In this talk I will review some of the recent applications of the replica theory to glasses.  I 
will 
firstly describe the basic assumptions and I will show that they can be considered as a precise 
reformulations of old ideas.  The relation of this approach with the mode-coupling theory will be 
shortly discussed.

I will present numerical simulations for binary mixtures.  The results of these simulations point 
toward the correctness of the replica approach to glasses.  I will describe the results of 
off-equilibrium simulations for large systems, in which the  aging dynamics is 
studied.
 \end{abstract}
 

In this talk I will review some of the theoretical progresses that have been done in our 
understanding 
of glasses.  They have been done mainly by comparing the results obtained for soluble models of 
generalized spin glasses with structural glasses assuming that the phase space of the two models 
are 
similar.

In the nutshell many of the ideas I am going to present are not new: they are already in the 
original papers of Gibbs and Di Marzio.  However the comparison of glasses and generalized spin 
glasses, introduced in ref.  \cite {KWT} allow us to put these ideas in a  sharper form and to 
test them in numerical (and eventually real) experiments.  In this talk I will not discuss the 
theoretical basis under which this scenario has been derived (i.e.  the mathematical tool needed to 
derive the results stated for the generalized spin glasses) but I will concentrate the attention of 
the physical picture.

The basic ideas are quite simple \cite{CUKU,FRAPA,PARE,kurparvir,crisomtap,I,remi}.  Let us 
consider 
a system of $N$ particles and let us assume that we can introduce a free energy functional 
$F[\rho]$ 
which depends on the density $\rho(x)$ and on the temperature.  We suppose that at sufficiently low 
temperature this functional has many minima (i.e.  the number of minima goes to infinity with the 
number ($N$) of particles).  Exactly at zero temperature these minima coincide with the mimima of 
the potential energy as function of the coordinates of the particles.  Let us label then by an 
index 
$\alpha$.  To each of them we can associate a free energy $F_\al$ and a free energy density $f_\al= 
F_\al/N$.  

In this low temperature region we suppose that the total free energy of the system can be well 
approximated by the sum of the contributions to the free energy of each particular minimum:
\be
Z\equiv \exp(-\beta N f_{S}) =\sum_\al \exp(-\beta N f_\al).
\ee

When the number of minima is very high, it is convenient to introduce the function $\cN(f,T,N)$ 
which is the density of minima whose free energy is equal to $f$.  With this notation we can write 
the previous formula as
\be
Z= \int df \exp (-\beta N f) \cN(f,T,N).
\ee
In the region where $\cN$ is exponentially large we can write
\be
\cN(f,T,N) \approx \exp(N\Sigma(f,T)),\label{CON}
\ee
where the function $\Sigma$ is called the complexity or the configurational entropy (it is the 
contribution to the entropy coming from the existence of an exponentially large number of locally 
stable configurations).

The relation (\ref{CON}) is valid in the region $f_m(T)<f<f_M(T)$.  The minimum possible value of 
the 
free energy  is given by $f_m(T)$. Outside this region we have that $\cN(f,T)=0$.  It 
all cases known $\Sigma(f_m(T),T)=0$, and the function $\Sigma$ is continuous at $f_m$.

For large values of $N$ we can write
\be
\exp(-N \beta f_{S}) \approx \int_{f_m}^{f_M} df \exp (-N(\beta f- \Sigma(f,T)).\label{SUM}
\ee
We can thus use the saddle point method and  approximate the 
integral  with the integrand evaluated at its maximum.
We find that
\be
\beta f_{S}=\min_f\Phi(f) \equiv \beta f^* - \Sigma(f^*,T),
\ee
where
\be
\Phi(f)\equiv\beta f - \Sigma(f,T).
\ee
(This formula is quite similar to the well known homologous formula for the free energy ,i.e.  
$\beta 
f=\min_{E} (\beta E -S(E))$, where $S(E)$ is the entropy density as function of the energy density.)

If we call $f^*$ the 
value of $f$ which minimize $\Phi(f)$. we have two possibilities:
\begin{itemize}
\item
The minimum is inside the interval and it can be found as solution 
of the equation $\beta=\partial \Sigma/\partial f$.  In this case we have
\be
\beta \Phi=\beta f^* - \Sigma^*, \ \ \ \Sigma^*=\Sigma(f^*,T).
\ee
The system may stay in one of the many possible minima.  The number of minima at which 
is convenient for the system to stay is $\exp(N \Sigma ^*)$ .  The entropy of the system is thus 
the 
sum of the entropy of a typical minimum and of $\Sigma^*$, which is the contribution to the entropy 
coming from the exponential large number of microscopical configurations.

\item
The minimum is at the extreme value of  the range of variability of 
$f$.  We have that $f^*=f_m$ and $\Phi=f_m$.  In this case the contribution of the complexity to 
the 
free energy is zero.  The different states who contribute the the free energy have a difference in 
free energy density which is of order $N^{-1}$ (a difference in total free energy of order 1).
Sometimes we indicate the fact that the free energy is 
dominated by a few different minima by say the the replica symmetry is spontaneously broken 
\cite{mpv,parisibook2}.
\end{itemize}

Form this point of view the behaviour of the system will crucially depend on the free energy 
landscape \cite{ACP}, i.d.  the function $\Sigma(f,T)$, the distance among the minima, the height 
of 
the barriers among them...  Although our final task should be to compute the properties of the free 
energy landscape from the microscopic form of the Hamiltonian, we can tentatively assume that the 
landscape of fragile glasses is similar to that of some soluble long range models in presence of 
quenched disorder \cite {KWT,ACP}.

I cannot discuss here in details the rationale for this hypothesis; it is also clear that it cannot 
be exact and some differences should be present among the predictions of the mean field 
approximation and the real world.  Here I will present the scenario coming from mean field, 
stressing some of the predictions that should have a wider range of validity and comparing them 
with 
numerical simulations for fragile glasses.

 We can distinguish a few temperature regions.
\begin{itemize}
\item For $T>T_f$ the only minimum of the free energy functional is given by a constant density.
The system is obviously in the fluid phase.
\item For $T_f>T>T_D$ there is an exponentially large number of 
minima with a non-constant density $\rho(x)$ \cite{kurparvir,babumez}.  There are values of the 
free 
energy density such that the complexity $\Sigma$ is different from zero, however the contribution 
coming from these minima is higher that the one coming from the liquid solution with constant 
$\rho(x)$.
\item The most interesting situation happens in the region where $T_D>T>T_K$.  In this region the
free energy is still given the fluid solution with constant $\rho$ and at the same time the free 
energy is also given by the sum over the non trivial minima \cite{I,remi}.

Although the thermodynamics is still given by the usual expressions of the liquid phase and final 
free energy is analytic at $T_D$, below this temperature the liquid phase correspond to a system 
which at each given moment may stay in one of the exponentially large number of minima.  It is 
extremely surprising that the free energy of the liquid can be written in this region as the sum of 
the contribution of the minima, according to formula (\ref{SUM}).  

The time to jump from one minimum to an other minimum is quite large: it is an activated process 
which is controlled by the height of the barriers which separate the different minima.  The 
correlation time will become very large below $T_D$ and for this region $T_D$ is called the 
dynamical transition point.  The correlation time (which should be proportional to the 
viscosity) should diverge at $T_{K}$.  The precise form of the this divergence is not well 
understood.
It is natural to suppose that we should get divergence of the form $\exp(A/(T-T_{K})^{\nu})$ for an 
appropriate value of $\nu$, whose reliable analytic computation is lacking \cite{KWT,PAK}.

The equilibrium complexity is different from zero (and it is a number of order 1) when the 
temperature is equal to $T_D$ and it decreases when the temperature decreases and it vanishes 
linearly at $T=T_K$.  At this temperature (the so called Kauzmann temperature) the entropy of a 
single minimum becomes to the total entropy and the contribution of the complexity to the total 
entropy vanishes.
\item
In the region where $T<T_K$ the free energy is dominated by the contribution of a few minima of the 
free energy having the lowest possible value.  Here the free energy is no more the analytic 
continuation of the free energy in the fluid phase.  A phase transition is present at $T_K$ and the 
specific heat is discontinuous here. 
\end{itemize}

\begin{figure}
\epsfxsize=250pt\epsffile{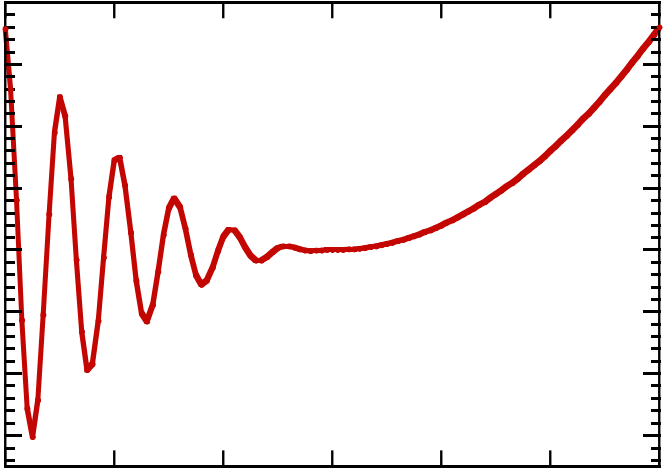}
\caption{The qualitative dependence of the free energy as function of the configuration space in 
the 
region relevant for the dynamical transition, i.e.  for $T<T_{D}$.  }
\end{figure}

The free energy landscape is quit usual; we can try do present the following pictorial 
interpretation, which is a rough simplification \cite{JL}.  At temperature higher than $T_{D}$ the 
system stays in a region of phase space which is quite flat and correspond of a minimum of the 
total 
free energy.  On the contrary below $T_{D}$ the phase space is similar to the one shown pictorial 
in 
fig.  1.  The region of maxima and minima is separated by the region without barriers by a large 
nearly flat region.  The minima in the region at the left are still present also when 
$T_{f}>T>T_{D}$, but they do not correspond to a global minimum.

At temperatures higher than $T_D$ the system at thermal equilibrium states in the right region.  
When the temperature reaches $T_D$ the system arrives in the flat region.  Here the potential is 
flat and this causes a more or less conventional Van Hove critical slowing down which is well 
described by the well known mode coupling theory \cite{vetro} (which is exact in the mean field 
approximation).  The mode coupling theory describes the critical slowing down which happens near 
$T_{D}$ \cite{BCKM}.

In the mean field approximation the height of the barriers separating the different minima is 
infinite and the temperature $T_D$ is sharply defined as the point where the correlation time 
diverge.  In the real world activated process (which are neglected in the mean field approximation 
and consequently in the mode coupling theory) have the effect of producing a finite (but large) 
correlation time also at $T_{D}$.  The precise meaning of the dynamical temperature beyond mean 
field approximation is discussed in details in \cite{FRAPA}

When the temperature is smaller that $T_D$ we must be more precise in describing the dynamics of 
the 
system.  Let us start from a very large system (of $N$ particles) at high temperature and let us 
gradually cool it. We find that it should go at equilibrium in the region with many minima.  
However 
coming from high free energy (from the right) it cannot enter in the region where are many maxima 
and minima if we wait a finite amount of time (the time to crosses the barriers diverges as $\exp 
(AN)$.  If we do not wait an exponentially large amount of time the system remains confined in the 
flat region.  In this case \cite{CUKU} the so called dynamical energy,
\be
E_D=\lim_{t \to \infty} \lim_{N \to \infty} E(t,N),
\ee
is higher that the equilibrium free energy.  The situation is described in fig.  2.

Below $T_D$ the system is trapped in metastable states when cooled.  The time needed to escape from 
these states diverges when $N$ goes to infinity in the mean field approach where activated 
processes 
are forbidden.  Of course the difference of the static and dynamic energy is an artifact of the 
mean 
field approximation if we take literarily the limit $t \to \infty$ in the previous equation because 
as matter of fact there are no metastable states with strictly infinite mean life.  However it 
correctly describe the situation on laboratory times, where metastable states are observed.

\begin{figure}
\epsfxsize=250pt\epsffile{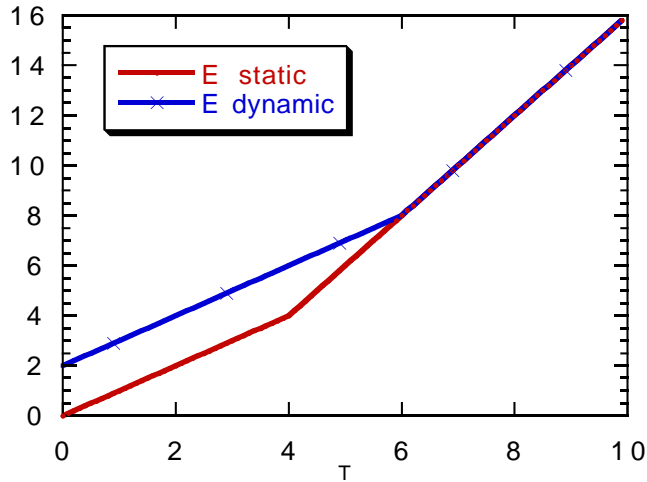}
\caption{The qualitative behaviour of the equilibrium and of the dynamical energy as function of 
the temperature.}
\end{figure}

In real systems, beyond the mean field approximation, the height of the barriers is finite also 
below $T_D$ and the mean life of the metastable states is finite, albeit very large.  Some 
mechanisms have been described who imply a divergence of the correlation time in real systems at 
the 
Kauzmann temperature \cite{KWT,FRAPA,PAK}.  The conventional glass temperature, i.e.  the 
temperature at which the microscopic correlation time becomes macroscopic (e.g.  of order of the 
minute) is between the two temperatures (i.e.  $T_D>T_G>T_K$).

It should be clear that in this framework the dynamical temperature $T_D$ is not so well defined 
and 
it correspond to a crossover region below it the dynamics is dominated by activated processes
\cite{FRAPA}.

In the mean field approximation there are very interesting phenomena that happen below $T_D$ when 
the system is cooled from the high temperature phase.  These phenomena are related to the fact that 
the system does not really go to an equilibrium configuration but wanders in the phase space never 
reaching equilibrium.  The phenomena are the following:
\begin{itemize}
\item
The energy approaches equilibrium slowly when the system is cooled from an high energy 
configuration.  In other words for large times we have
\be
E(t,T)=E_{D}(T)+B(T) t^{-\lambda(T)},
\ee
where the exponent $\la(T)$ does not vanish linearly at zero temperature as it should happens for 
an 
activated process.
\item
Aging is present, i.e.  the correlation functions and the response functions in the region of large 
time do depend on the story of the system \cite{B,POLI,FM}.
\item
In the region where aging is present the hypothesis at the basis of the fluctuation dissipation 
theorem are no more valid.  New generalized relations are satisfied \cite{FRARIE,MPRR}, which 
replace the fluctuation dissipation theorem.
\item
In the region where the diffusion constant is zero, a new phenomenon, logarithmic diffusion, is 
present.
\end{itemize}
All these phenomena are well known also in the case of spinodal decomposition but have a more 
general validity.

Here I will illustrate this phenomena in the case of a monatomic fluid (I will consider the case of 
a binary mixture in order to avoid crystalization).  The particle interacts with a soft $x^{-12}$ 
potential. 
 Half of the particles 
are of type $A$, half of type $B$ and the interaction among the particles is given by the 
Hamiltonian:
\begin{equation}
H=\sum_{{i<k}} \left(\frac{(\si(i)+\si(k)}{|{\bf x}_{i}-{\bf x}_{k}|}\right)^{12},\label{HAMI}
\label{HAMILTONIAN}
\end{equation}
where the radius ($\si$) depends on the type of particles.

The model is well studied in the past \cite{HANSEN1,HANSEN2,HANSEN3,PAAGE,KOB}.  It is known that 
the choice $\si_{B}/\si_{A}=1.2$ strongly inhibits crystallisation and the system goes into a 
glassy 
phase when it is cooled.  Using the same conventions of the previous investigators we consider 
particles of average radius $1$, i.e.  $ ({\si_{A}^{3}+ 2 (\si_{A}+\si_{B})^{3}+\si_{B}^{3})/4}=1 
$.  
The system goes into the glass phase at $\Gamma
\approx 1.45$, where we follow the standard notation: $\Gamma\equiv\beta^{1/4}$.

For reason of space I will concentrate on the properties of diffusion \cite{PAAGE3}.  Let me 
firstly 
discuss the equilibrium situation and later I will discuss the non-equilibrium case.

A typical quantity that can be studied in a glass is the diffusion.  We can define the average 
distance squared done by a particle in time $t$ as
\be
\Delta(t)=\lan (x_i(t)-x_i(0))^2 \ran\label{DELTA}
\ee
where the label $i$ may take values from 1 to $N$.  The diffusion constant $D$ is defined as
\be
\Delta(t) \approx D t
\ee
for large $t$.  

In the following we will assume (and it is a rather good approximation) that in the glass phase 
$\Delta(t)$ goes to a constant ($\Delta^*$) at large $t$, so that the diffusion constant $D$ is 
zero.

We can also introduce the response to an external force.  Let us consider the following time 
dependent Hamiltonian
\be
H=H_0 +\eps(t) F\cdot x_i,\label{TDH}
\ee
where $H_0$ is the original unperturbed Hamiltonian and $F$ is a vector of length $d$, where $d$ is 
the dimension of the space (usually 3).  Both here and in eq.  (\ref{DELTA}) the value of index $i$ 
(i.e the particle we have chosen) is arbitrary and the result does not depend on $i$.

We consider the case where
\be
\eps(t)= 0 \for t<0,\ \ \ \ \eps(t)=\eps \for t>0,
\ee
$\eps$ being a small number.  The response function is defined as
\be
\lan F \cdot x_i(t) \ran = \eps R(t) +\lan F \cdot x_i(t) \ran |_{\eps=0}.
\ee
The usual fluctuation dissipation theorem tell us that
\be
\beta \Delta(t) =   R(t),
\ee
 In the large $t$ limit we obtain the 
celebrated Einstein relation among diffusion and viscosity.
\begin{figure}[htbp]
\epsfxsize=250pt\epsffile{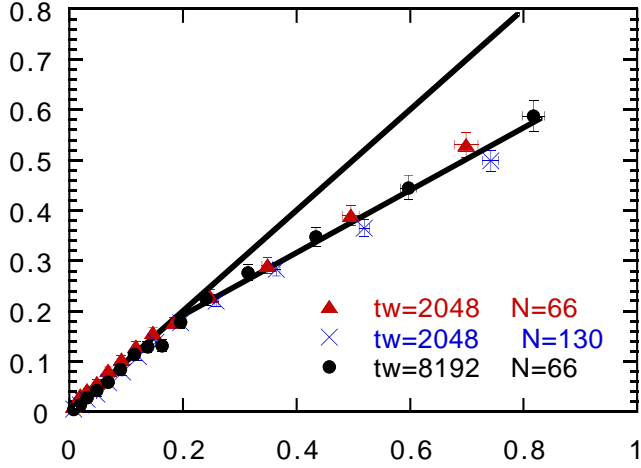}
\caption{ $R$ versus $\beta\Delta$ at $\Gamma=1.6$ for 
$t_{w}=8192$ and $t_{w}=2048$ at $N=66$ and for $t_{w}=2048$ at $N=130$.  The two straight lines 
have slope 1 and .62 respectively.}
\end{figure}

Let us consider an off-equilibrium situation.  Here the results depend on the history of the 
system.  
We a very simple history.  The system is in the liquid phase for $t<0$, at $t=0$ it is cooled very 
fast at the final temperature in the glassy phase.  Here time translational invariance is no more 
present.  We generalize the previous definitions introducing two times quantities:
\be
\Delta(t,t_w)=\lan(x_i(t+t_w)-x_i(t_w))^2\ran.
\ee
The response function is defined by
\be
\eps R(t,t_w)=\lan x_i(t+t_w)\cdot F\ran-\lan F \cdot x_i(t+t_w) \ran {\Biggr |}_{\eps=0}
\ee
where the statistical expectation values are computed with the time dependent Hamiltonian 
(\ref{TDH}), where $\eps(t)$ given by
\be
\eps(t)= 0 \for t<t_{w},\ \ \ \ \eps(t)=\eps \for t>t_{w}
\ee
We obviously have that 
\be
\lim_{t_w \to \infty} \Delta(t,t_w)=\Delta(t)
\ee
and in this way we recover the equilibrium limit.

The equilibrium limit is relevant when $t_w>>t$.  The aging regime is present when both $t$ and 
$t_{w}$ are both large.  In the simpler case (i.e.  naive aging) the interesting behaviour happens 
when $t=O(t_w)$ (i.e.  the ratio $t/t_w$ is finite) and $t_w$ is large.  In this regime we 
find that
\be
\Delta (t,t_w)\approx {\bf \Delta}\apa{ t \over t_w}\cpa .
\ee
In other words if we measure something on a time scale $t$ comparable to $t_w$, we always see a 
non-equilibrium behaviour.

The generalized fluctuation dissipation theorem tells us that, if we plot at a given $t_w$ $\Delta$ 
versus $R$, the plot has a limit when $t_w$ goes to infinity (as an effect of aging) and we can 
define a function $X(\Delta)$ \cite{CUKU} as
\be
\frac{\partial R}{\partial \Delta}= \beta X(\Delta).
\ee

The function $X$ can be constructed for any pairs of observable and its value should not depend on 
the observables.  More precisely, if we take two other quantities satisfying the fluctuation 
dissipation theorem, we should have that at equal values of $s$ the function $X$ takes the same 
value of $X(\Delta)$.  The fluctuation dissipation theorem is violated in the same way for all the 
observables at the same time.

It has been conjectured that this dynamically defined function $X$ is related to an similar 
function 
which can be defined by measurements done only at equilibrium \cite{CUKU}.  A precise presentation 
of this conjecture would take too much time.  A numerical verification of its correctness in short 
range spin glasses can be found in \cite{MPRR}.

The simplest non trivial form of the function $X(\Delta ) $ is (in analogy to what happens in
generalized spin glasses)
\bea
X(\Delta ) =1 \for \Delta<\Delta^{*}\\
X(\Delta ) =m(T) \for \Delta>\Delta^{*}
\eea
where $m(T)$ is approximately a linear function of the temperature taking the value one at 
$T=T_{D}$.

\begin{figure}
\epsfxsize=250pt\epsffile{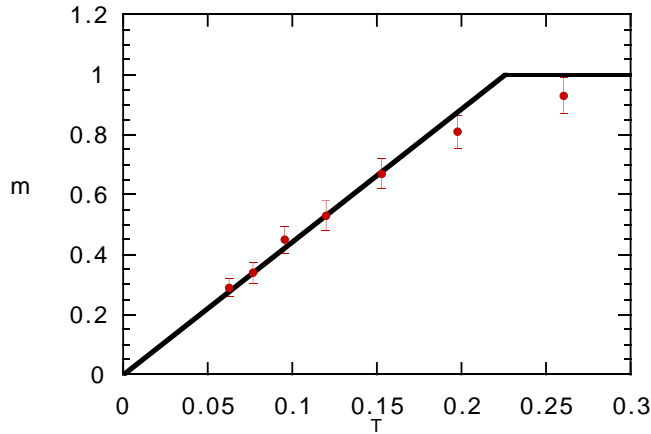}
\caption{ The quantity $m\equiv	{\partial R / \partial \beta	\Delta}$ as $t_{w}=2048$ as
function of	the	temperature. The straight line is
the	prediction of the approximation	$m(T)=T/T_{c}$.}
\end{figure}

In other words if we plot the response $R$ as function of $\beta \Delta$ we should find piece-linear
graph with slope 1 for at small $\Delta$ and slope $m$ for large $\Delta$.
The result of numerical simulations in the glassy phase are shown in fig.  3 (for details see ref.
\cite{PAAGE3}).

Naive aging would imply that in the large time region (both times large) $\Delta(t,t_w)$ is a 
function of only the ration $t/t_{w}$.  If we do some reasonable assumption naive aging implies  
that 
$\Delta(t,t_w)$ behaves as $A\ln(t/t_{w})+B$ when $t$ is large \cite{RUOCCO}.

This phenomenon is well known in Ising spin systems.  It amount to say that if we cool at zero 
temperature a spin system starting from an high energy configuration, the average numer of times a 
spin is reversed increases as the logarithm of the time.  This off-equilibrium logarithmic 
diffusion 
seems to be present in glasses.  For example numerical simulations of a binary mixture are very 
well 
in agreement with this prediction as can be seen in fig.  5.
\begin{figure}
\epsfxsize=250pt\epsffile{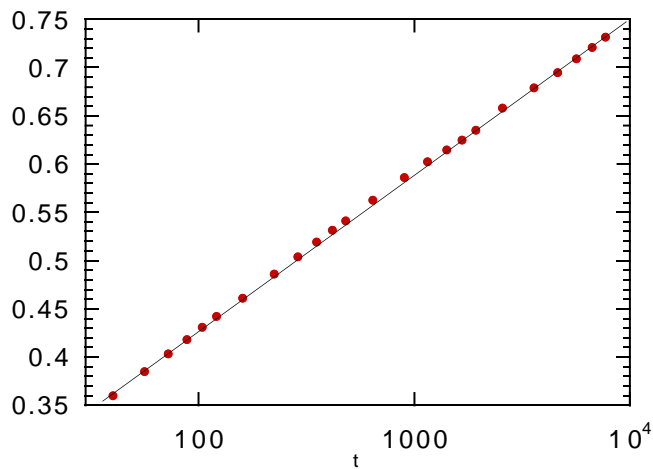}
\caption{The average distance squared from the initial configuration (i.e.  $\Delta(t,1)$) as 
function
of
the time in a logarithmic scale  for  $\Gamma=1.6$.}
\end{figure}

The results of simulations are in very good agreement with the theoretical expectations based on 
our 
knowledge extracted from the mean field theory for generalized spin glass models.  The 
approximation 
$m(T)=T/T_{D}$ seems to work with an embarrassing precision.  We can conclude that the ideas 
developed for generalized spin glasses have a much wider range of application than the models from 
which they have been extracted.  It likely that they reflect quite general properties of the phase 
space and therefore they can be applied in cases which are very different from the original ones.  
In a recent work \cite{FRAPA} some thermodynamic predictions have been obtained for the behaviour 
of 
glassy systems, like the present ones, under the assumption that they obey the laws derived for 
generalized spin glasses.  The present work confirms that assumption.

The most urgent theoretical task would be now to develop an analytic theory for glasses in the low 
temperature region from which one could compute the function $m(T)$.  This goal should not be out 
of 
reach: a first step in this direction can be found in \cite{MEPA}.

\end{document}